\documentclass[10pt,letterpaper]{article}
\usepackage{opex3}
\usepackage{threeparttable}
\usepackage{cite}

\begin{document}

\title{On the dispersion management \\
       of fluorite whispering-gallery mode \\
       resonators for Kerr optical frequency \\
        comb generation in the telecom \\
       and mid-infrared range}

\author{Guoping Lin$^*$ and Yanne K. Chembo}
\address{FEMTO-ST Institute [CNRS UMR6174], Optics Department, \\
         15B Avenue des Montboucons, 25030 Besan\c con cedex, France.\\
         }
\email{$^*$guoping.lin@femto-st.fr} 



\begin{abstract}
Optical whispering gallery mode (WGM) resonators have been very attracting platforms for versatile Kerr frequency comb generations. We report a systematic study on the material dispersion of various optical materials that are capable of supporting quality factors above $10^9$. Using an analytical approximation of WGM resonant frequencies in disk resonators,
we investigate the effect of the geometry and transverse mode order on the total group-velocity dispersion ($GVD$). 
We demonstrate that the major radii and the radial mode indices play an important role in tailoring the $GVD$ of WGM resonators. 
In particular, our study shows that in WGM disk-resonators, the polar families of modes have very similar $GVD$, while the radial families of modes feature dispersion values that can differ by up to several orders of magnitude.
The effect of these giant dispersion shifts are experimentally evidenced in Kerr comb generation with magnesium fluoride.    
From a more general perspective, this critical feature enables to push the zero-dispersion wavelength of fluorite crystals towards the mid-infrared (mid-IR) range, thereby allowing for efficient Kerr comb generation in that spectral range. 
We show that barium fluoride is the most interesting crystal in this regard, due to its zero dispersion wavelength ($ZDW$) at $1.93~\rm{\mu m}$ and an optimal dispersion profile in the mid-IR regime. 
We expect our results to facilitate the design of different platforms for Kerr frequency comb generations in both telecommunication and mid-IR spectral ranges.  
\end{abstract}

\ocis{ 
 (140.4780) Optical resonators;
 (260.1180) Crystal optics;
 (260.2030) Dispersion;
 (190.4380) Nonlinear optics, four-wave mixing. 
} 


\section{Introduction}
Optical frequency combs have been very useful scientific and technological tools in a wide range of applications such as optical clocks, precision spectroscopy, low phase noise microwave generation and optical communications. The general approach for producing frequency combs is based on mode-locking in lasers. An alternative approach has been recently demonstrated in a silica monolithic toroidal microresonator~\cite{Haye2007Optical}, where cavity enhanced four-wave mixing (FWM) occurs. As a result, this type of comb using the third-order susceptibility in resonators is usually referred to as Kerr optical frequency comb. The emergence of such combs provides a compact and energy efficient solution. Therefore, extensive amount of works in both theories and experiments have been devoted to this field. For example, the understanding of the comb formation is now very elaborate, and involves both coupled-mode and spatiotemporal  models~\cite{Chembo2010Spectrum,Chembo2010Modal,Matsko2011Mode,Chembo2013Spatiotemporal,Coen2013Modeling,Godey2014Stability}; various applications have also been experimentally explored ranging from low phase noise microwave oscillations~\cite{Li2012Low,Savchenkov2013Stabilization}, ultra-short pulse generation~\cite{Saha2013Modelocking,Herr2014Temporal}, and optical communications~\cite{Pfeifle2014Coherent}. 

Compared with Kerr comb generation platforms based on chip-scale waveguide racetrack resonators~\cite{Razzari2010CMOS}, the mechanically polished disk resonators have a much wider range of material choices and can feature comb generation with very low pump power in the sub-mW level~\cite{Savchenkov2013Stabilization}, which results from the ultra-high quality ($Q$) factors and small mode volumes in these WGM resonators. Although Kerr frequency combs in fluoride WGM resonators have been demonstrated in the telecommunication window and recently at $2.5~\rm{\mu m}$ in the mid-IR regime~\cite{Wang2013Mid}, there is still not a systematic study on the optimal choice of materials and the sizes of resonators. Moreover, anomalous $GVD$ is generally desirable for comb generation and the spanning of Kerr combs is also strongly dependent on the dispersion profile. Recently, an analytical approximation for the dispersion of WGMs in the disk geometry has been derived by two different groups, showing a good agreement with  finite-element simulations and high-precision free spectral range (FSR) measurements~\cite{Breunig2013Whispering,Demchenko2013Analytical}. It enables an accurate study of the impact of geometry on the $GVD$. On the other hand, barium fluoride (BaF$_2$) with a good transmission range covering the whole mid-IR window has been recently shown to be capable of supporting $Q$ factors above $10^9$~\cite{Lin2014Barium}, and this result broadens the choice of crystalline candidates for Kerr comb generation.

In this article, we report the study of material dispersion for different IR and mid-IR optical materials using Sellmeier equations. We point out that fluoride crystalline materials generally have significantly flatter material dispersion profile than other materials. A systematic and theoretical investigation of the impact of resonator geometry and WGM indices on the $GVD$ is carried out, showing that the major radii and the radial WGM indices are key parameters to consider for $GVD$ tailoring in WGM resonators. The impact of geometry on $GVD$ is also experimentally evidenced in an experiment showing how primary Kerr combs with significantly different multiplicity can be generated with the same MgF$_2$ resonator, but pumped in WGMs with different radial orders. We also investigate theoretically MgF$_2$ and BaF$_2$ platforms for potential broadband comb generations in the telecom and mid-IR regimes.

\section{In search of the optimal materials for Kerr comb generation}

\begin{figure}[t]
\centering\includegraphics[width=9cm]{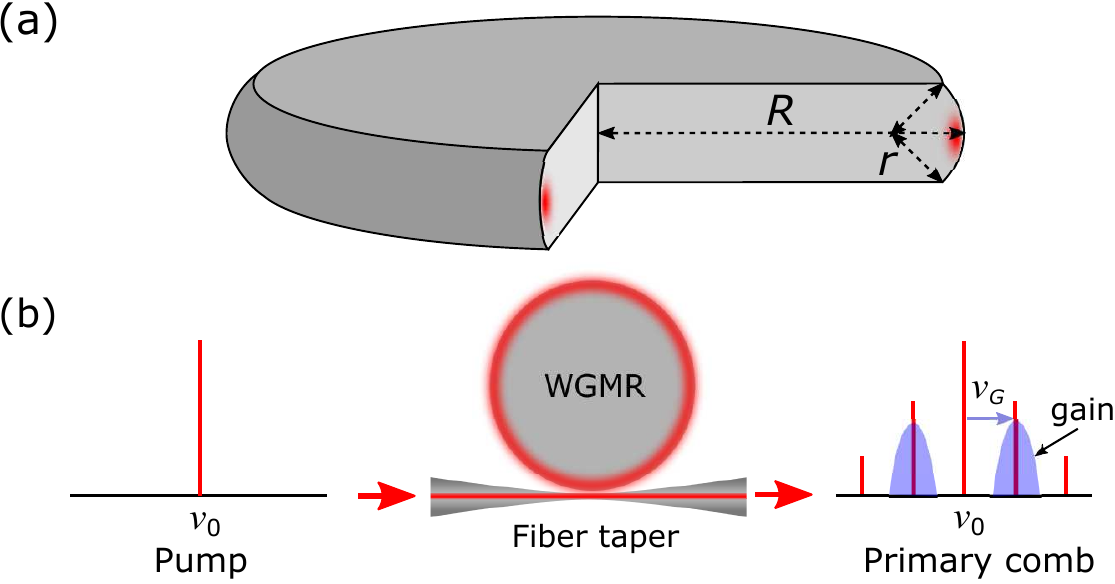}
\caption{ (a) Geometry of the disk resonator. (b) Schematic of Kerr frequency comb generations in a fiber taper coupled resonator setup. WGMR: WGM resonator. $v_G$ is the primary parametric gain peak spacing from the pump frequency $v_0$.}
\label{fig:schematic}
\end{figure}

As far as Kerr frequency comb generation is concerned, the primary step has been the search of the optimal materials. Although the on-chip ring resonator as comb generation platforms is much more compact, mechanically polished high $Q$ WGM crystalline resonators are still of great interest, due to the flexible choice of materials and ultra-high $Q$ factors leading to power efficient operation. 

In these axi-symmetric disk resonators, the rim part can be treated as a toroidal boundary with the main radius $R$ and the curvature radius $r$, as illustrated in Fig.~\ref{fig:schematic} (a). Figure~\ref{fig:schematic} (b) gives a typical schematic drawing of Kerr comb generation. A single frequency continuous-wave (cw) pump laser can be coupled into the WGMR through a fiber taper or a prism~\cite{Gorodetsky1999Optical}. Under certain conditions that have been analyzed in 
refs.~\cite{Chembo2010Spectrum,Chembo2010Modal,Godey2014Stability}, the built-up intracavity light leads to the modulation instability (MI) gain with offset $v_G$ from the pump frequency $v_0$. The parametric FWM then creates the first two sidebands where WGMs with single or multiple free spectral range ($FSR$) relative to the pump mode meet the MI gain. Subsequent parametric process creates a comb with multiple-FSR spacing, generally referred to as the \textit{primary comb}. Depending on the dispersion and pump condition, broadband Kerr combs with single-FSR spacing and thousands of frequency lines can be further produced. As Kerr comb arises from MI gain, it thus prefers an anomalous dispersion profile similar to MI gain in optical fibers. In addition, the MI theory in Kerr comb generation predicts that the offset $v_G$ of the parametric gain peak frequency scales with the dispersion following $v_G \propto 1/\sqrt{GVD}$~\cite{Chembo2010Spectrum,Chembo2010Modal,Godey2014Stability}. Hence, weak dispersion is generally favorable for broadband Kerr comb generation. It should be noted that the primary Kerr comb shown in Fig.~\ref{fig:schematic} (b) also yields rolls (or Turing patterns) in the spatial domain~\cite{Godey2014Stability, Coi13Azimuthal}.

At this date, Kerr optical frequency combs have been reported on WGM resonators made of fused silica, magnesium fluoride, calcium fluoride (CaF$_2$) and sapphire~\cite{PhysRevLett.101.093902,Savchenkov2011Kerr,Grudinin2012Frequency,Wang2013Mid,Grudinin2013Impact,Herr2014Temporal,Ilchenko2014Generation}, which are all capable of featuring $Q$ factors above $10^9$. Recently, such high $Q$ factors have also been demonstrated on barium fluoride~\cite{Lin2014Barium}. 
We thereby take into account this material with others for the investigation of the material dispersion ($GVD_M$) using their Sellmeier equations~\cite{Milam1977Nonlinear}. The $GVD_M$ is usually characterized by group delay dispersion (in ps/km/nm): 
\begin{equation}
GVD_m = - \frac{\lambda}{c} \frac{\partial^2 n (\lambda)}{\partial \lambda^2}
\label{Equ:GVD}
\end{equation}
where $\lambda$ is the wavelength, $c$ is the speed of light in vacuum, $n(\lambda)$ is the refractive index.

\begin{figure}[t]
\centering\includegraphics[width=9cm]{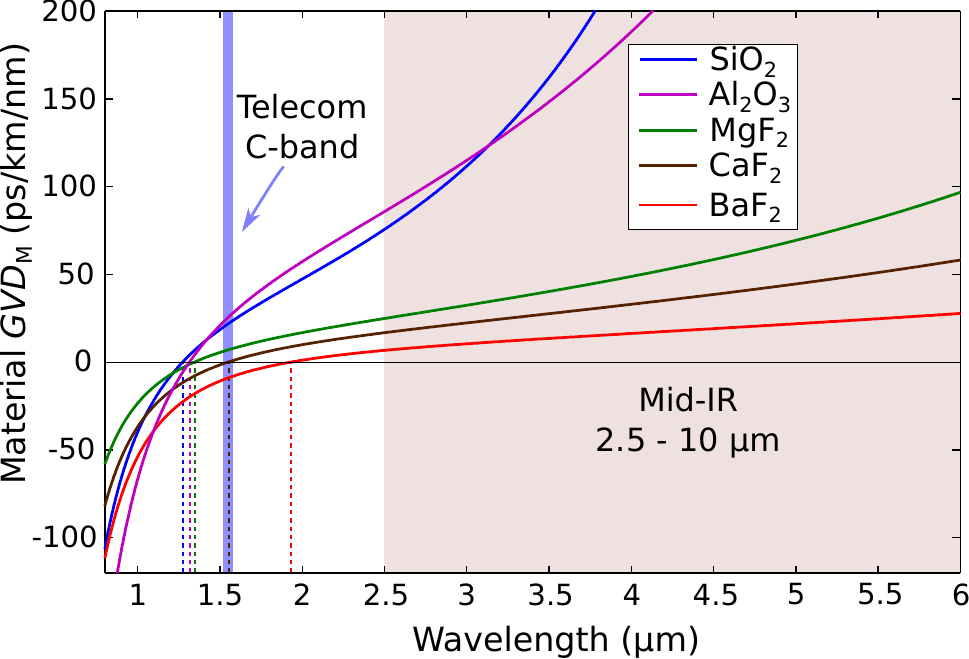}
\caption{Comparison of material dispersion for different optical materials: fused silica (SiO$_2$), sapphire, MgF$_2$, CaF$_2$ and BaF$_2$. Dashed lines designate the corresponding $ZDWs$: $1.27~\rm{\mu m}$, $1.31~\rm{\mu m}$ $1.34~\rm{\mu m}$, $1.55~\rm{\mu m}$ and $1.93~\rm{\mu m}$, respectively. The vertical blue and brown zones highlight the telecommunication C-band and mid-IR windows. Note that the $GVD_M$ profile of MgF$_2$ investigated here is derived for the ordinary ray. The extraordinary one has very similar profile with a slightly different $ZDW$ at $1.57~\rm{\mu m}$.}
\label{fig:Dmaterial}
\end{figure}

Figure~\ref{fig:Dmaterial} displays the corresponding material $GVD_M$ as a function of wavelength from 0.8 to 6~$\rm{\mu m}$. It should be noted that the dispersion is anomalous when $GVD$ is greater than zero. A broadband comb generation will require a small $GVD$ value in this regime. On the other hand, a $GVD$ that is too large will also make Kerr comb generation difficult, as the corresponding parametric gain could be too close to the pump frequency and do not meet any WGMs (the phase matching condition will therefore never be fulfilled). One can also see that fluoride materials have flatter dispersion profiles than the others. 
The corresponding $ZDW$ of these materials are listed in Table~\ref{tab:material}. Also shown are some optical characteristics of these materials~\cite{Weber2002Handbook}. 

\begin{table} [b]
\centering\caption {Characteristics of bulk fused silica, Sapphire, MgF$_2$, CaF$_2$, BaF$_2$ at $T\approx300~K$ for Kerr comb generation\textsuperscript{a}} \label{tab:material}
\begin{threeparttable}
 	\begin{tabular}{l| p{1.7cm}| p{1cm}| p{1.6cm}| p{1.9cm}| p{2.4cm}}
\hline
Name & Transmission ($\rm{\mu m}$) & $\lambda_{0}$ ($\mu m$) & Refractive index at $\lambda_{0}$  & $(dL/L)/dT$ ($10^{-6}$/K) &  $dn/dT$ ($10^{-6}$/K)\\
\hline
Fused silica &  0.2-4.0 & 1.27 & 1.4473 & 0.55 & 9.6\\
Sapphire &  0.19-5.2 & 1.313 & 1.7502 & 6.65(a)/7.15(c) & 13.6(o)/14.7(e)\\
MgF$_2$ & 0.13-7.7 & 1.34 & 1.3717 & 9.4(a)/13.6(c) & 0.9(o)/0.3(e)\\
CaF$_2$ & 0.12-10 & 1.55 & 1.4260 &  18.9 & -11.5\\
BaF$_2$ & 0.14-13 & 1.93 & 1.4648 & 18.4 & -16.2\\
\hline
	\end{tabular}
     \begin{tablenotes}
       \item[a] $(dL/L)/dT$ along the a and c axes are denoted. o: ordinary ray; e: extroordinary ray. $dn/dT$ for fluoride crystals and sapphire are given at $1.15\mu m$ and $0.589\mu m$, respectively. $dn/dT$ for fused silica at $1.13\mu m$ is taken from manufacturing specifications.
     \end{tablenotes}
\end{threeparttable}
\end{table}

One can note that fused silica, quartz and MgF$_2$ have anomalous $GVD_M$ around the telecom wavelength, while all the fluoride crystals are transparent well into the mid-IR regime and also feature anomalous $GVD_M$. Among them, BaF$_2$ is characterized by a flat high transmission  covering the whole mid-IR regime ($2.5 - 10~\rm{\mu m}$). Moreover, the reported $2~\rm{nm}$ surface roughness in BaF$_2$~\cite{Lin2014Barium}, also confirms its potential for reaching ultra-high $Q$ factors that are only limited by the material absorption across the whole mid-IR regime. Also shown in the table are their thermal expansion coefficients $(dL/L)/dT$ and thermo-optic coefficients $dn/dT$. These parameters are particularly important for WGM applications. The self-thermal locking method can be used for WGM pumping for microlasers~\cite{Lin2012Thermal} and combs~\cite{Haye2007Optical}. However, it is only applicable for the material that possesses the same signs of these two parameters, such as MgF$_2$, fused silica and Sapphire. On the other hand, the others could benefit from self-thermal compensation like athermal glasses. Hence, they are less sensitive to the environment temperature fluctuation, due to the fact that the frequency variation $dv$ of a WGM is determined by $(dv/v)/dT \approx (dn/n)/dT -(dL/L)/dT$~\cite{Lin2014Continuous}. Considering another useful locking technique, the self-injection locking method using Rayleigh scattering induced feedback is applicable in both materials~\cite{Liang2010Whispering,Savchenkov2013Stabilization}.

\section{Geometry dispersion effects of WGMs in disk resonators}

It is known that the geometry dispersion can have a strong impact in the total dispersion, such as that of the silicon channel waveguide~\cite{Turner2006Tailored}. Similarly, it also plays an important role in resonator structures. As light in WGM resonators is confined under the equatorial boundary of the disk, the size and the curvature of the rim are expected to affect the total dispersion value. A recent study also shows that the dispersion of a spherical resonator can be tailored by a coating layer~\cite{Ristic2014Tailoring}. Here, we focus on monolithic resonators without any coating to preserve their high $Q$ factors that are only limited by the absorption of the host material. 

A simple analytical approximation of the eigenfrequencies of  WGMs derived by two groups~\cite{Breunig2013Whispering,Demchenko2013Analytical} will be used in this work.  The  WGM resonant wavelength in a toroid can be expressed as follows:
\begin{equation}
\frac{2 \pi n(\lambda) R}{\lambda} = m - \alpha_q \left( \frac{m}{2}\right)^{1/3} + \left( p+ \frac{1}{2} \right) \left( \frac{R}{r} \right)^{1/2}
\label{Equ:WGMPosition}
\end{equation}
where $n(\lambda)$ is determined by the Sellmeier equation of the material, $m, q, p$  are the mode numbers in the azimuthal, radial and polar directions, $\alpha_q$ is the negative $q$th root of the Airy function. The index $p$ is similar to that of WGMs in spheres and is related to two momentum numbers $l$ and $m$ by $p = l - m$, indicating the $p + 1$ antinodes of the field distribution in the polar direction which can be identified through excitation mapping~\cite{Lin2010Excitation}. Other methods from spectroscopy and emission patterns for identifying WGMs have also been demonstrated~\cite{Schiller1991High,Schunk2014Identifying}. In a typical mm-size disk resonator, $m$ usually ranges from $10^3$ to $10^5$ for optical resonances.

It should be noted that Kerr combs have been generated using two different sets of WGMs. The first one is the family of WGMs with spacing of one FSR, $v_{\rm{FSR}}=v_{m} -v_{m-1}$ and the other one is a group of polar WGMs with $v_{p} -v_{p-1}$ spacing~\cite{Savchenkov2011Kerr}, where $v = c / \lambda$ is the resonant frequency. Here, we focus on the Kerr combs generated using the longitudinal FSR, and which can feature hundreds to thousands of comb lines, interesting for many spectroscopy applications. We therefore analyze the total dispersion of these longitudinal FSR spaced WGMs. The dispersion of WGMs underlying Kerr comb generation is characterized by the variation in the FSR, $\Delta v_{\rm{FSR}} = (v_{m+1} -v_{m})-(v_{m} -v_{m-1})$. 
This value can be derived by numerically solving Eq.~\ref{Equ:WGMPosition} with the underlying Sellmeier equations.
To convert $\Delta v_{\rm{FSR}}$ into the widely used dispersion parameter $GVD$ in units of ps/nm/km, we applied the following approximated expression~\cite{Arcizet2009Optical}:
\begin{equation}
GVD \approx 4 \pi^2 \frac{n(\lambda)^3 R^2}{c^2 \lambda^2} \Delta v_{\rm{FSR}} \, .
\label{Equ:Conversion}
\end{equation}

\begin{figure}[t]
\centering\includegraphics[width=12cm]{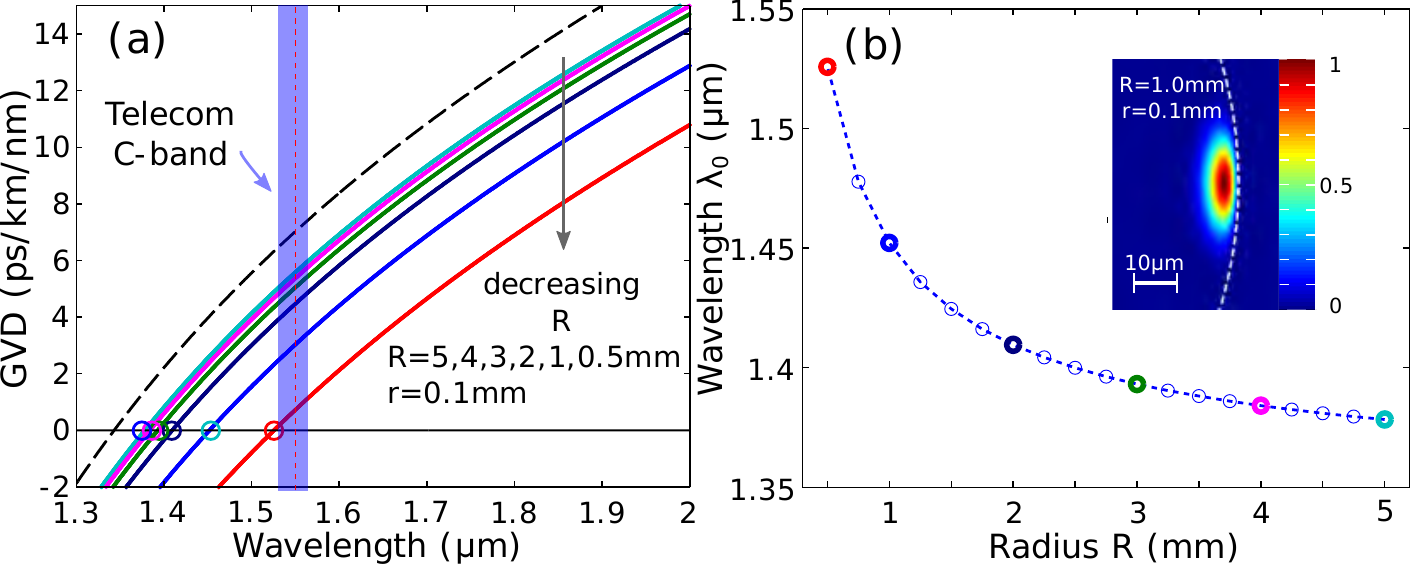}
\caption{(a) Calculated geometry dispersion influence on the total $GVD$ for a MgF$_2$ disk with different $R$ and $r=0.1$~mm. Dashed line: material $GVD_M$ curve. (b) Calculated zero total dispersion wavelengths $\lambda_0$ for $R$ from $0.5$~mm to $5$~mm. Inset: distribution of the electric field amplitude for a fundamental WGMs with $R=1$~mm and $r=0.1$~mm. Note: $\lambda_0$ is the resonant wavelength that has its total dispersion closest to $0$.}
\label{fig:MgF2WGMSize}
\end{figure}


Here, we will study the impact of geometry dispersion by assuming WGM resonators made of MgF$_2$, which has a small $GVD$ over a large range when compared with fused silica and quartz. We first consider the case of the fundamental WGM family with $q = 1$ and $p = 0$ and discuss the size impact of $R$ and $r$. As one can see in Eq.~(\ref{Equ:WGMPosition}), the second term on the right side is much larger than the third term due to the large value of $m$ in mm-size resonators. We thus study only the impact of the main radius $R$ on the total dispersion, because the effect of $r$ is similar to the effect of the polar number $p$ that will be introduced later. Figure~\ref{fig:MgF2WGMSize} (a) shows the calculated total $GVD$ for different radii $R$ values from $0.5$~mm to $5$~mm around the zero dispersion regime. The dashed line presents the material $GVD_M$. 

One can note that smaller $R$ shifts the zero $GVD$ further into the longer wavelengths, which results from a normal geometry $GVD$ adding on to the material $GVD_M$. The total $GVD$ profile compared to the material one is eventually pulled down. Similar effects have also been reported~ \cite{Grudinin2013Impact,Li2012Sideband}. The wavelengths ($\lambda_0$) of the WGMs closest to zero $GVD$ as a function of the radii $R$ is shown in Fig.~\ref{fig:MgF2WGMSize} (b). The inset provides a calculated electric field distribution of a fundamental WGM resonating at $\lambda_0$ in a MgF$_2$ disk with $R = 1.0$~mm and $r=0.1$~mm using an analytical expression given in~\cite{Breunig2013Whispering}. From this figure, one can conclude that small anomalous $GVD$ can be engineered by carefully choosing the major radius of the resonator. This is a key step for designing a Kerr frequency comb setup. For example, the $GVD$ for the fundamental modes in a $R=0.5$~mm MgF$_2$ resonator at $1.55~\rm{\mu m}$ is reduced to $0.72$~ps/km/nm, which is about one tenth of the original material $GVD$ ($7.0$~ps/km/nm). Careful engineering of this parameter, the $GVD$ reduction can be even larger.
\begin{figure}[t]
\centering\includegraphics[width=12cm]{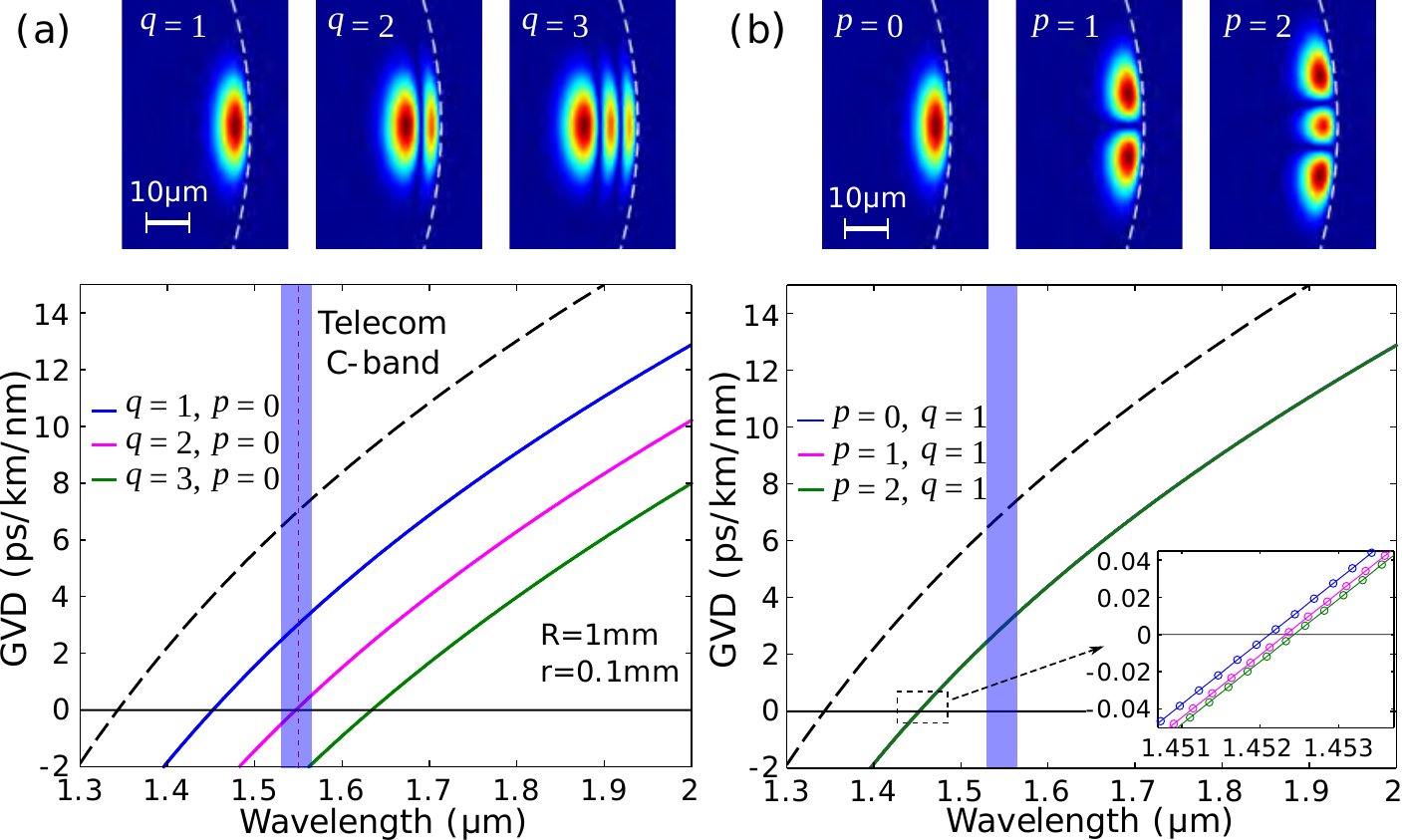}
\caption{Calculated total $GVD$ for different transverse WGMs in a MgF$_2$ disk with $R=1$~mm and $r=0.1$~mm. Top: Electric field amplitude distribution of WGMs; Bottom: $GVD$ as a function of the resonant wavelength. For WGMs with transverse indices of (a) $q = 1, 2, 3$, $p = 0$ and (b) $p = 0, 1, 2$, $q = 1$. Dashed line: material $GVD_M$ curve. }
\label{fig:MgF2WGMindices}
\end{figure}

The ability of the $GVD$ tailoring in a WGM resonator after fabrication is also critical. Thereby, we have also investigated the effect of different transverse mode orders on the total $GVD$. Figure~\ref{fig:MgF2WGMindices} shows the corresponding calculation for a MgF$_2$ resonator with $R = 1.0$~mm and $r=0.1$~mm. The black dashed lines marks the material dispersion itself. The upper images are the corresponding calculated field distribution patterns. The impact of WGMs with different $q$ indices on the total $GVD$ is presented in Fig.~\ref{fig:MgF2WGMindices} (a). 
On the other hand, the mode index $p$ plays a less interesting role in tailoring the total $GVD$ as shown in Fig.~\ref{fig:MgF2WGMindices} (b). The inset is a zoom in of the $GVD$ profile around the $ZDW$. The red dashed line highlights the telecom wavelength of $1.55~\rm{\mu m}$. The fundamental radial mode with $q = 1$ possesses a $GVD$ of $3.0$~ps/km/nm, which is much larger than the previous value for the fundamental mode in a resonator with $R=0.5$~mm. However, one can note that for the second radial order WGMs with $q=2$, the $GVD$ has been further reduced to $~0.086$~ps/km/nm, that is only $~1/80$ of the material $GVD_M$ and $1/35$ of that for $q=1$. In this case, Kerr comb generations can be significantly different, dependent on which radial order mode is excited. Comparing Fig.~\ref{fig:MgF2WGMindices} with Fig.~\ref{fig:MgF2WGMSize}, one can see that the management of $GVD$ by choosing WGMs with different $q$ indices is more remarkable. Besides, we also find that the change of $GVD$ due to the $q$ index variation is much larger in the resonator with a smaller $R$.

\section{Experimental observation of significant different primary combs in a single MgF$_2$ resonator}

To experimental verify that the geometry dispersion plays an important role in WGM resonators, we have carried out experiments on a disk resonator made of MgF$_2$. The resonator was fabricated by carefully polishing the rim of a disk preform with a radius of $6$~mm and a thickness of $1$~mm. It features an intrinsic $Q$ factor of $10^9$. For a well polished resonator, WGMs with lower order mode indices can possess the same $Q$ factors and cannot be distinguished through $Q$ factor measurements. It has been experimentally shown that one can choose the fiber taper to excite WGMs with the radial indices from $q=1$ to $q=11$ simultaneously~\cite{Li2012Sideband}. Hence, a fiber taper was used to couple the light into and out from the resonator as illustrated in Fig.~\ref{fig:schematic} (b). The throughput of the fiber taper was then monitored with a photodetector and a high-resolution spectrum analyzer (APEX 2440B), respectively. The tunable cw pump laser with a sub-KHz spectral linewidth at $1552$~nm was used. It was then thermally locked to excite different WGMs~\cite{Haye2007Optical,Lin2012Thermal}. Depending on the pump condition, both primary Kerr combs and full developed Kerr combs were observed. The value of single FSR spacing ($v_{\rm{FSR}}$) can thus be measured.

\begin{figure}[h]
\centering\includegraphics[width=13cm]{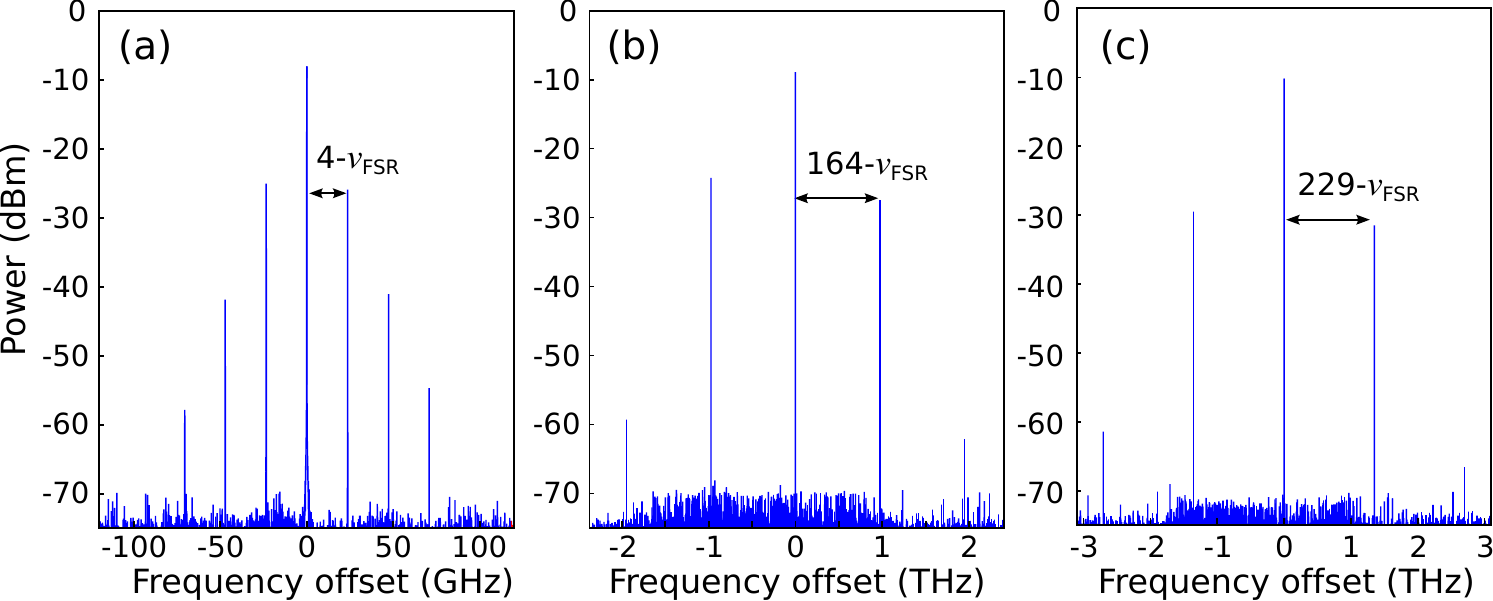}
\caption{Experimental observation of three primary combs showing very different spacing of (a) $4$-$v_{\rm{FSR}}$, (b) $164$-$v_{\rm{FSR}}$ and (c) $229$-$v_{\rm{FSR}}$. 
Three different WGMs were pumped in the same resonator, and the large difference of $GVD$ explains the large difference of primary comb multiplicity. }\label{fig:MgF2Combs}
\end{figure}

The giant shift in dispersion predicted by our calculations should therefore allow for the possibility to generate primary combs with significantly different multiplicities, by pumping different radial modes in the same resonator. 
In particular, one should be able to observe very different primary comb spacing in the same resonator with a fixed incident pump power, especially in the case of fiber taper coupling where many different radial order WGMs can be excited at the same time. Figure~\ref{fig:MgF2Combs} present three experimentally obtained primary combs in the same MgF$_2$ resonator when three different WGMs were pumped. The spacing of $23.7$~GHz , $971.9$~GHz and $1351.1$~GHz corresponds to $4$-$v_{\rm{FSR}}$, $164$-$v_{\rm{FSR}}$ and $229$-$v_{\rm{FSR}}$, respectively. The corresponding ratios between the largest spacing and the smallest one is about $50$. Although the detuning between the pump and the resonance can also change the spacing of the primary comb, the difference is usually limited due to the requirement of on-resonance pumping. In our case, we usually see a ratio of less than $2$ when changing the detuning within one mode with a fixed incident pump power. Therefore, we believe that the geometry dispersion differences cause this observation, which is expected as previously discussed. 


\section{Towards the mid-IR regime}

As shown in Table~\ref{tab:material}, BaF$_2$ has a larger zero material dispersion wavelength ($\lambda_0$) of $1.93~\rm{\mu m}$, compared with other materials studied here. In addition, the $GVD_M$ of BaF$_2$ in the anomalous regime also grows much slower. This is an important signature indicating that it could be very favorable for the broadband comb generation in the mid-IR regime. As shown in Fig.~\ref{fig:Dmaterial}, $2.5 \rm{\mu m}$ marks the first Kerr frequency comb generated in the mid-IR regime~\cite{Wang2013Mid}, where a MgF$_2$ WGM resonator was utilized. It also indicates that comb generation is feasible for a material $GVD_M$ of $25~\rm{ps/km/nm}$. It should be noted that BaF$_2$ has a nonlinear refractive index coefficient ($n_2$, in $10^{-16}\rm{cm^2/W}$) of $2.85$, which is larger than $1.90$ of CaF$_2$ and $0.92$ of MgF$_2$~\cite{Milam1977Nonlinear}. If we assume a BaF$_2$ resonator with the same $Q$ factor and the same geometry dispersion, one can thus expect a Kerr comb generation capability at $\lambda = 5.5 \rm{\mu m}$. Considering that the demonstrated Kerr comb at $2.5 \rm{\mu m}$ in MgF$_2$ has a very high phase stability~\cite{Wang2013Mid}, the capability of generating such combs further into the wavelength above $5 \rm{\mu m}$ in BaF$_2$ is thus highly desirable.

\begin{figure}[h]
\centering\includegraphics[width=12cm]{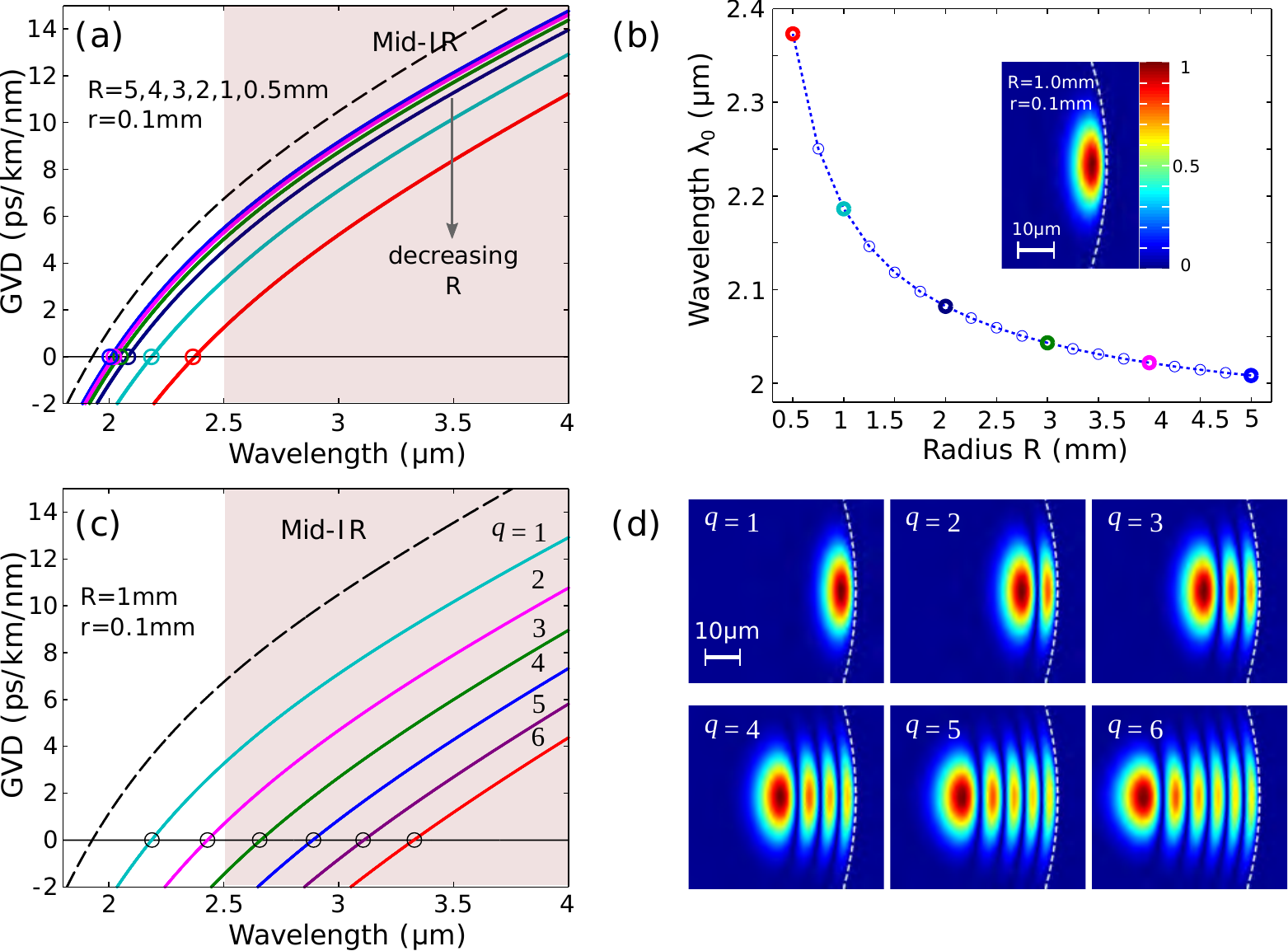}
\caption{ (a) Calculated geometry dispersion influence on the total $GVD$ for a disk BaF$_2$ with different $R$ and $r=0.1$~mm. Dashed line: material $GVD_M$ curve.. 
(b) Calculated zero total dispersion wavelengths $\lambda_0$ for $R$ from $0.5$~mm to $5$~mm. Inset: distribution of the electric field amplitude for a fundamental WGM with $R=1$~mm and $r=0.1$~mm. 
(c) Calculated radial mode order influence on the total $GVD$ for a disk BaF$_2$ with $R=1$~mm and $r=0.1$~mm. WGM indices: $q = 1, 2, 3, 4, 5, 6$ and $p = 0$. (d) The corresponding electric field amplitude distribution of WGMs.}
\label{fig:BaF2WGM}
\end{figure}

Here, we present the corresponding $GVD$ calculation on WGMs made of BaF$_2$ for Kerr comb generations towards mid-IR. Figure~\ref{fig:BaF2WGM} (a) shows the $GVD$ for the fundamental WGMs in resonators with different $R$ and $r=0.1$~mm. As BaF$_2$ has a material $GVD_M$ at $1.93~\rm{\mu m}$, it is thereby feasible to push the zero dispersion WGM resonance wavelength $\lambda_0$ close to $2.5~\rm{\mu m}$, which can potentially facilitate broadband comb generations into this regime. The corresponding $\lambda_0$ as a function of the main radii $R$ are shown in Fig.~\ref{fig:BaF2WGM} (b). The inset gives the field distribution of a fundamental mode at about $2.2~\rm{\mu m}$ in a resonator with $R=1$~mm and $r=0.1$~mm. 

Figure~\ref{fig:BaF2WGM} (c) presents the $GVD$ as a function of WGM resonance wavelengths for different radial order modes in a BaF$_2$ resonator with $R=1$~mm and $r=0.1$~mm. The corresponding field distribution is shown in Fig.~\ref{fig:BaF2WGM} (d). As one can see, $GVD$ values at $3.5~\rm{\mu m}$ for different radial WGMs are anomalous and reduced with the increase of $q$ indices. In the case of prism coupling setup, one can change the coupling angle to selectively excite WGMs with high $q$ values. The corresponding wavelengths and indices of those modes closest to the zero total dispersion in Fig.~\ref{fig:BaF2WGM} (c) are listed in Table~\ref{tab:WGMs}. For $q=1$ and $q=2$, the change in $ZDW$ of about $240$~nm is observed. $ZDW$ beyond $3~\rm{\mu m}$ is feasible in higher $q$ order modes. 

\begin{table} [h]
\centering\caption {Indices of WGMs and zero total dispersion wavelength for MgF$_2$ with $R=1$~mm and $r=0.1$~mm } \label{tab:WGMs}
\begin{tabular}{p{2cm} p{0.8cm} p{0.5 cm} p{0.5cm} |p{2cm} p{0.8cm} p{0.5 cm} p{0.5 cm}}
\hline
Wavelength ($\lambda_0(WGM)$) & $m$ & $q$ & $p$ & Wavelength ($\lambda_0(WGM)$) & $m$ & $q$	 & $p$ \\
\hline
2.1863 & 4176 & 1 & 0 & 2.8877 & 3100 & 4 & 0\\
2.4290 & 3733 & 2 & 0 & 3.1100 & 2860 & 5 & 0\\
2.6608 & 3386 & 3 & 0 & 3.3272 & 2656 & 6 & 0\\

\hline
\end{tabular}
\end{table}

\section{Conclusion}

In conclusion, we have investigated the material dispersion profiles of different optical materials that support WGMs with $Q$ factors above $10^9$. We have studied the effect of geometry dispersion on the total $GVD$ of WGM disk-resonators. This analysis has enabled us to demonstrate that the major radii of WGM resonators and the radial order indices are key parameters for dispersion tailoring. In particular, we have shown that the dispersion shift induced by selecting different radial families of modes in the same resonator allows for the generation of versatile combs.
Barium fluoride could benefit from this feature and it offers the best dispersion profile in the mid-IR spectral range.
We expect this work to facilitate the efficient generation of Kerr optical frequency combs in the telecom and mid-IR regimes.  

\section*{Acknowledgments} 
{The authors acknowledge financial support from the European Research Council (ERC) through the projects NextPhase and Versyt.
They also acknowledge financial support from the \textit{Centre National d'Etudes Spatiales} (CNES) through the project SHYRO, from the \textit{R\'egion de Franche-Comt\'e}, and from the Labex ACTION.}

\end{document}